\documentclass[10pt]{iopart}
\usepackage{graphicx}
\usepackage{latexsym}
\newcommand {\be}{\begin{equation}}
\newcommand {\ee}{\end{equation}}

\begin{document}

\paper[]{Off-equilibrium Langevin dynamics of the discrete nonlinear
Schr\"odinger chain}
\date{\today}

\author{S Iubini$^{1,2}$, S Lepri$^{1}$, R Livi$^{2}$ and A Politi$^{3}$\
}
\address{$^{1}$ Istituto dei Sistemi Complessi, Consiglio Nazionale
delle Ricerche, via Madonna del Piano 10, I-50019 Sesto Fiorentino, Italy
}
\address{$^{2}$ Dipartimento di Fisica e Astronomia - CSDC, Universit\`a di Firenze
and INFN Sezione di Firenze, 
via G. Sansone 1 I-50019, Sesto Fiorentino, Italy 
}
\address{$^{3}$  Institute for Complex Systems and Mathematical Biology \& SUPA
University of Aberdeen, Aberdeen AB24 3UE, United Kingdom 
}
\ead{stefano.iubini@fi.isc.cnr.it}

\begin{abstract}
We introduce suitable Langevin thermostats  which are able to control
both the temperature and the chemical potential of a one-dimensional
lattice of nonlinear Schr\"odinger oscillators. The resulting 
non-equilibrium stationary states are then investigated in the limit of
low temperatures and large particle densities, where the dynamics can 
be mapped onto that of a coupled-rotor chain with an external torque. 
As a result, an effective kinetic definition of temperature can be 
introduced and compared with the general microcanonical (global) definition.

\end{abstract}

\submitto{Journal of Statistical Mechanics: theory and experiment}

\section{Introduction}
Understanding transport properties in open many--particle systems
is one of the main goals of contemporary nonequilibrium statistical mechanics.
The ultimate goal is to find the statistical measure for
stationary out-of-equilibrium conditions. In fact, this would allow
evaluating the fluctuations of the relevant macroscopic observables (such
as the currents) and, possibly, deriving the corresponding transport
equations, without any \textit{ad hoc} statistical assumption.
In view of the many technical difficulties that one typically encounters along this
path, it is convenient to start investigating simple models, like chains of 
nonlinear oscillators \cite{LLP03,DHARREV}.
A particularly interesting system is the Discrete NonLinear Schr\"odinger (DNLS) equation
\cite{Eilbeck1985,Kevrekidis} that  has important applications  in many domains of
physics. A classical example is electronic transport in biomolecules
\cite{Scott2003}, while in optics or acoustics it describes the
propagation of nonlinear waves in a layered photonic or phononic media
\cite{Kosevich02,Hennig99}. With reference to cold atomic gases, the DNLS equation 
provides an approximate semiclassical description of bosons trapped in periodic optical 
lattices (for a recent survey see \cite{Franzosi2011} and references therein).

This system is rather interesting since the presence of two conserved quantities 
(energy and number of particles) naturally requires arguing about coupled transport, 
in the sense of ordinary linear irreversible thermodynamics. In fact, in spite of the 
many studies of heat conduction in oscillator chains~\cite{LLP03,DHARREV,Delfini07b,BBO06},
coupled transport processes have been scarcely 
investigated~\cite{Gillan85,Mejia2001,Larralde03,Casati2009,Saito2010}.
DNLS studies have been so far mostly focused on its dynamical properties such as the emergence of
breather solutions~\cite{Flach2008}. The first analysis of the equilibrium properties of 
the continuous nonlinear Schr\"odinger equation has been presented in \cite{Lebowitz1988}, while
a similar analysis of the DNLS was developed more recently in \cite{Rasmussen2000}. 
Extensions to a wider class of DNLS-like equations can be found in~\cite{Johansson2006}. 
On the other hand, the analysis of the DNLS nonequilibrium properties 
is still in its early days \cite{Basko2011,Iubini2012}.

The statistical analysis of any system of physical interest requires a proper 
modelling of the interaction with an external reservoir. The reservoir is expected 
to exchange energy and particles with the system until a steady state is reached,
characterized by the expected temperature and chemical potential.
One of the most powerful approaches is based on the introduction of suitable
stochastic differential equations such as for the Langevin thermostats 
that are typically considered in the study of oscillator chains
(see, e.g., \cite{LLP03,DHARREV}).
In models such as the DNLS equation, this option is less straightforward, 
because of its non-separable structure. In fact, the only previous study made
used of Monte Carlo thermostats \cite{Iubini2012}. In this paper we bridge
the gap by augmenting the Hamiltonian equations with a suitable nonlinear 
damping and a stochastic term. To our knowledge, this is the first such 
scheme to be proposed in the literature,
at least in the present context, although one should mention
\cite{Christiansen1997a}, where the evolution of a DNLS system 
has been discussed in the presence of small nonlinear damping and a 
multiplicative noise.

This general Langevin scheme is first used to verify the convergence
to equilibrium and then to investigate transport properties 
in two limit cases, low temperatures and large particle densities, 
where the DNLS model reduces to a chain of coupled oscillators with
internal forces. Such a relationship with translationally invariant
models helps to understand that the origin of the normal 
transport observed in DNLS chains~\cite{Iubini2012} is more subtle
than one might have thought. 
In fact, translationally invariant systems typically exhibit 
diverging transport coefficients \cite{LLP03}: only in models 
characterized by phase-like variables (such as the XY model)
transport is normal because of the occasional scattering
of the phonons with phase-jumps across the energy barriers \cite{LLP03}.

Moreover, we find that the chemical potential, which is associated with norm conservation,
is equivalent to the rotation frequency of the single rotors and
the corresponding force that must be exerted by the external Langevin reservoir for
its thermalization is an additional constant torque.
Finally, the possibility to map the DNLS equation onto a standard chain of coupled (phase)
oscillators allows deriving a local microscopic definition of the temperature, based on
their kinetic energy. This quite interesting since, so far, the only available definition
of temperature \cite{Franzosi2011} is both nonlocal and rather convoluted.

The paper is organized as follows. In section \ref{sec:dnls} we
introduce the model and recall its basic properties.
In section \ref{sec:lang} we present the Langevin equations and discuss
the equilibrium setup, as well as the case of two external reservoirs 
at different temperatures and chemical potentials.
In section~\ref{sec:har} we discuss the low-temperature and large mass-density limits
showing that the model can be mapped onto a chain of coupled (nonlinear) rotors.
A numerical test of the kinetic definition of the temperature 
is provided in section \ref{sec:sim}, while section \ref{sec:end} is devoted to a final
discussion of the achievements and to a presentation of future perspectives.
Finally, the two appendices contain an alternative derivation of the generalized Langevin
equations and the details of the derivation of the low-temperature Hamiltonian.

\section{The DNLS model at equilibrium}
\label{sec:dnls}
\begin{figure}[ht]
\begin{center}
\includegraphics[width=0.7\textwidth,clip]{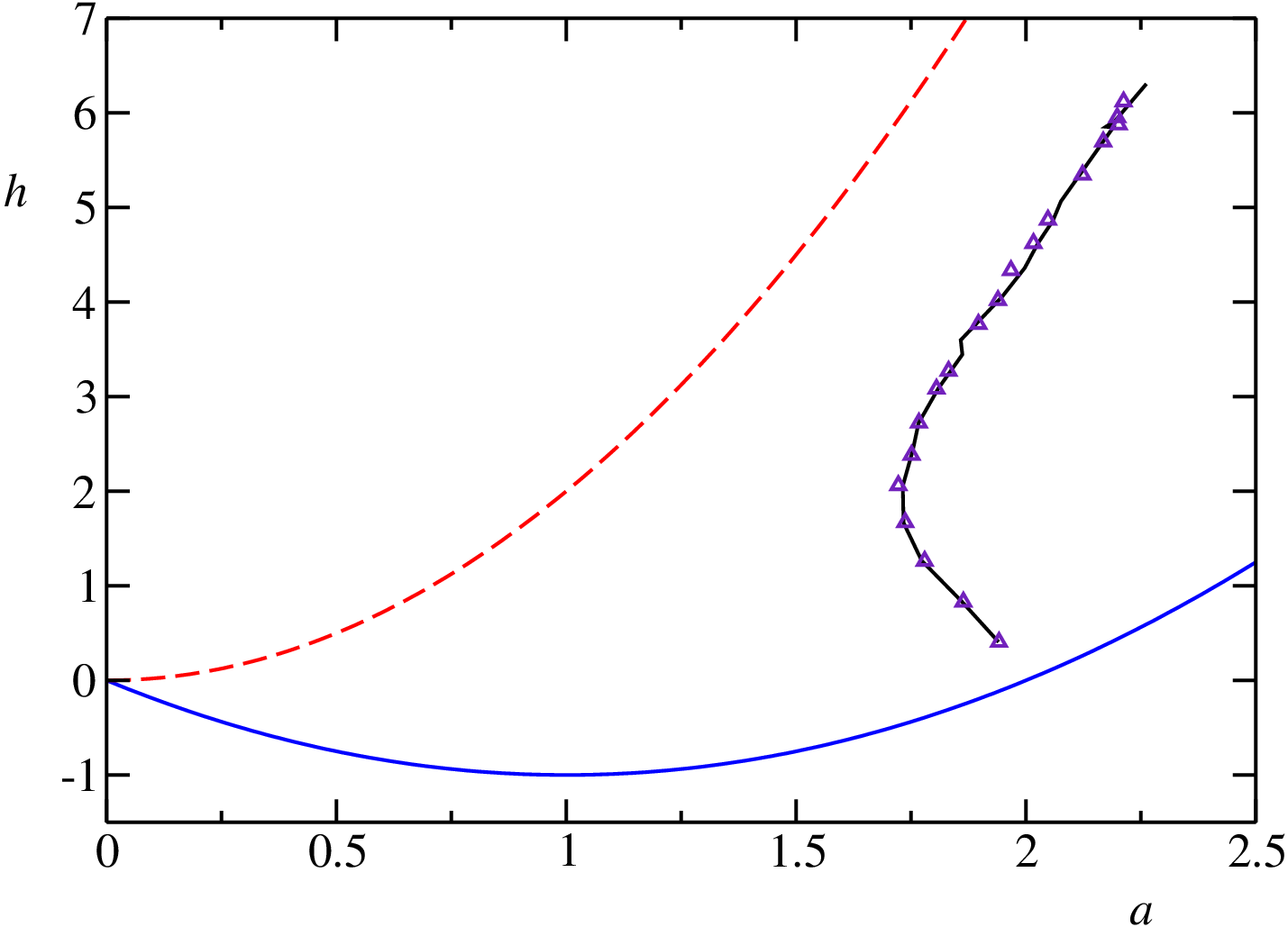}
\caption{Equilibrium phase diagram of the DNLS model in the $(a,h)$ plane. 
The positive temperature region lies between the ground state (solid blue line) and the 
infinite temperature isothermal (dashed red line). The line at constant
chemical potential $\mu=2$ has been obtained,  for  $ 0.5 \leq T \leq 10$, 
making use of the Monte--Carlo
stochastic thermostats \cite{Iubini2012} (purple triangles) and the Langevin  scheme
(solid black line).
}
\label{f:ph_diag}
\end{center}
\end{figure}

In this section we introduce the model and summarize its basic properties.
The Hamiltonian of a one-dimensional DNLS chain on a lattice with $N$ sites
can be written as (in suitable adimensional units)
\begin{equation}
 H= \sum_{n=1}^N \left( |z_n|^4+z_n^*z_{n+1}+z_nz_{n+1}^* \right) \quad ,
\label {Hz}
 \end{equation}
where $z_n$, $iz^*_n$ are complex, canonical coordinates, and $|z_n|^2$ 
can be interpreted as the {\sl number of particles},  or, equivalently, 
the {\sl mass} at site $n$.
The sign of the quartic term is assumed to be positive, i.e. we consider the case of
repulsive on--site interaction, while the sign of the hopping
term is irrelevant, due to the symmetry associated with the canonical (gauge)
transformation $z_n \to z_ne^{i\pi n}$. 
The corresponding equations of motion, $\dot z_n= -\partial H/\partial {i z^*_n}$, read
as
\begin{equation}
\label{uno}
i \dot {z}_n = -2 |z_n|^2z_n - z_{n+1}-z_{n-1} \quad .
\label{eqmot}
\end{equation}
For later reference, it is also convenient to introduce the real-valued canonical 
coordinates
\begin{equation}
 p_n=\frac{z_n+z^*_n}{\sqrt{2}},\qquad
 q_n= \frac{z_n-z^*_n}{ \sqrt{2} i}\quad ,
\end{equation}
which allow rewriting the equations of motion (\ref{eqmot}) as
\begin{eqnarray}
  \dot p_n&=&-(p_n^2+q_n^2)q_n-q_{n+1}-q_{n-1}\\
 \dot q_n&=&(p_n^2+q_n^2)p_n+p_{n+1}+p_{n-1}  \nonumber \quad .
\end{eqnarray}
An important property of the DNLS dynamics is the presence of a second conserved
quantity (besides energy), namely, the total mass
\begin{equation}
A=\sum_{n=1}^{N} |z_n|^2 = \frac{1}{2}\sum_{n=1}^{N}(p_n^2+q_n^2) \quad ,
\label{norm}
\end{equation} 
As a result, the equilibrium phase-diagram is two-dimensional, as it involves 
the energy density $h=H/N$ and the mass density $a=A/N$ (within a microcanonical 
description) or, equivalently, the temperature $T$ and the chemical 
potential $\mu$ (within a grand-canonical description). The first reconstruction
of the equilibrium phase-diagram was reported in \cite{Rasmussen2000} with reference
to the grand-canonical ensemble, with the help of transfer integral techniques. 
It is schematically reproduced in figure~\ref{f:ph_diag}, where the lower solid line
\begin{equation}\label {t0line}
h = a^2-2a  \quad,
\end{equation}
corresponds to the ground state ($T=0$) 
\begin{equation}\label{groundstate1}
z_n = \sqrt{a} \,\, {\rm e}^{i[\omega t+\pi n]}\quad,
\end{equation}
for different values of the chemical potential $\mu = \omega = 2(a - 1)$ \cite{Rasmussen2000}.
Positive-temperature states lie above such a curve, up to the dashed line
\begin{equation}
h = 2a^2 \,  
\label{tinfline}
\end{equation}
which corresponds, to infinite-temperature (and $\mu = -\infty$), 
characterized by random phases and an exponential distribution of the amplitudes.

Finally, above the $T=\infty$ line, one finds the so-called negative-temperature 
states~\cite{Rasmussen2000}. In this region, the dynamics of the DNLS equation is 
characterized by long-lived localized excitations (discrete breathers) \cite{Rumpf2004}.
We refer to \cite{Iubini2013} and to the literature cited therein for a discussion
of such peculiar states, whose properties have not yet been fully clarified.

\section{Langevin thermostats}
\label{sec:lang}

The equilibrium properties of the positive-temperature states have been previously explored
with the help of Monte--Carlo (MC) thermostats \cite{Iubini2012}, under the assumption 
of the existence of a grand-canonical statistical measure. As a result, it is for instance
possible to reconstruct the states characterized by constant chemical potential (see the
triangles in figure~\ref{f:ph_diag}, which correspond to $\mu=2$).

In this section we present an alternative approach, based on Langevin thermostats \cite{LLP03}.
It allows for a more rigorous mathematical formulation and a more direct physical interpretation.
In separable Hamiltonian systems (i.e., those composed of a kinetic and a potential energy)
interaction with a Langevin bath simply amounts to 
modify the momentum equation, by adding a linear dissipation
term $-\gamma p_n$, accompanied by a white-noise fluctuation, whose amplitude
determines the temperature value. This simple scheme does not work
for the DNLS. In fact, one can easily check that in the limit of vanishing
fluctuations (which, supposedly correspond to the zero-temperature limit) this dissipative dynamics 
converges to a fixed point, that does not correspond to the ground state, which, as 
mentioned in the previous section, is a time-periodic solution~\cite{Rasmussen2000}.

The problem can be overcome by adopting the following scheme,
\begin{equation}
\hspace{-1.cm} i \dot z_n= (1+i \gamma)\left[-2|z_n|^2 z_n  -z_{n+1}-z_{n-1} \right]  
 +i\gamma \mu z_n+ \sqrt{\gamma T} \, \xi_n(t) \quad ,
\label{ollac}
\end{equation} 
where $\xi_n(t)=\xi'_n+i\xi''_n$ is a complex, Gaussian, white random noise 
with unit variance and $\gamma$ is the coupling strength with the bath.
In practice, the above equation, which is basically a stochastic, discrete, 
complex Ginzburg-Landau equation, corresponds to a series of thermostats
all operating at temperature $T$ and chemical potential $\mu$, attached
to each lattice site. As required for a meaningful reservoir, the dissipative 
term vanishes along the  ground state evolution, $z_n = \sqrt{a} \exp{[i(\omega t +\pi n)]}$. 

At variance with MC schemes, one can show (with the help of suitable assumptions and approximations)
that eq.~(\ref{ollac}) describes the coupling with an ensemble 
of complex linear oscillators (see the derivation in Appendix 1). 
Additional physical insight is gained by rewriting Eq.~(\ref{ollac}), in terms of the $p_n$, 
$q_n$ variables,
\begin{eqnarray}\label{olla}
 \dot p_n&=&-\frac{\partial H}{\partial q_n}-\gamma\frac{\partial H_\mu}{\partial p_n} 
 +\sqrt{2\gamma T} \xi_n'(t)\\
 \dot q_n&=&\frac{\partial H}{\partial p_n} -\gamma\frac{\partial H_\mu}{\partial q_n} 
 +\sqrt{2\gamma T} \xi_n''(t) \nonumber \quad ,
\end{eqnarray}
where and $H_\mu$ is the effective Hamiltonian  $H_\mu=H-\mu A$. 
In the absence of thermal noise, the deterministic components of the thermostat,
being gradient terms, would drive the system towards a state characterized by a minimal 
$H_\mu$. The presence of the symplectic forces allows navigating across the microstates
characterized by the given $H_\mu$-value. These considerations suggest that this is the proper
way to define a dissipation scheme in the DNLS case. Actually, the reason why $H_\mu$ is considered 
instead of $H$ is the presence of two conservation laws: the minimum of the energy depends 
on the mass density $a$. In order to ensure the convergence to the proper state, the
term $-\mu A$ must be added to the effective Hamiltonian. The additional presence of the 
fluctuations completes the definition of the generalized Langevin equation, which represents
a proper stochastic reservoir for the DNLS equation with temperature $T$ and chemical potential $\mu$.
From (\ref{olla}) one can also check that the related Fokker-Planck
equation admits as a stationary solution the expected equilibrium grandcanonical distribution 
$\exp\{-\beta(H-\mu A)\}$, with $\beta=1/T$.
This setup can be straightforwardly implemented to investigate non--equilibrium settings, by 
assuming that the single reservoirs operate at different temperatures/chemical potentials. 

In the following we will focus on a typical setup adopted in the study of stationary
nonequilibrium regimes: only the first ($z_1$) and the last ($z_N$) lattice variables 
interact with the reservoirs.
This means that, assuming fixed boundary conditions (i.e. $z_0 = z_{N+1} =0$), the evolution on the 
leftmost site is ruled by the equation
\begin{eqnarray}
\label{olla2}
\hspace{-1.cm} 
\dot p_1&=&-(p_1^2+q_1^2)q_1-q_{2}-\gamma\left[(p_1^2+q_1^2)p_1+p_2 -\mu_L p_1 \right] 
 +\sqrt{2\gamma T_L} \xi'_1 \\ 
\hspace{-1.cm} \dot q_1&=&(p_1^2+q_1^2)p_1+p_{2} -\gamma\left[(p_1^2+q_1^2)q_1+q_2-\mu_L q_1\right] 
 +\sqrt{2\gamma T_L} \xi''_1 \nonumber  \quad ,
\end{eqnarray}
where $T_L$ and $\mu_L$ denote the temperature and the chemical potential
of the left reservoir, respectively. Analogous equations hold for the right reservoir, which
acts on the site $n=N$, where the temperature is $T_R$, the chemical potential is $\mu_R$,
and the coupling strength is set again equal to $\gamma$ for simplicity.
The rest of the chain follows the Hamiltonian evolution~(\ref{eqmot}).
\begin{figure}[ht]
\begin{center}
\includegraphics[width=0.7\textwidth,clip]{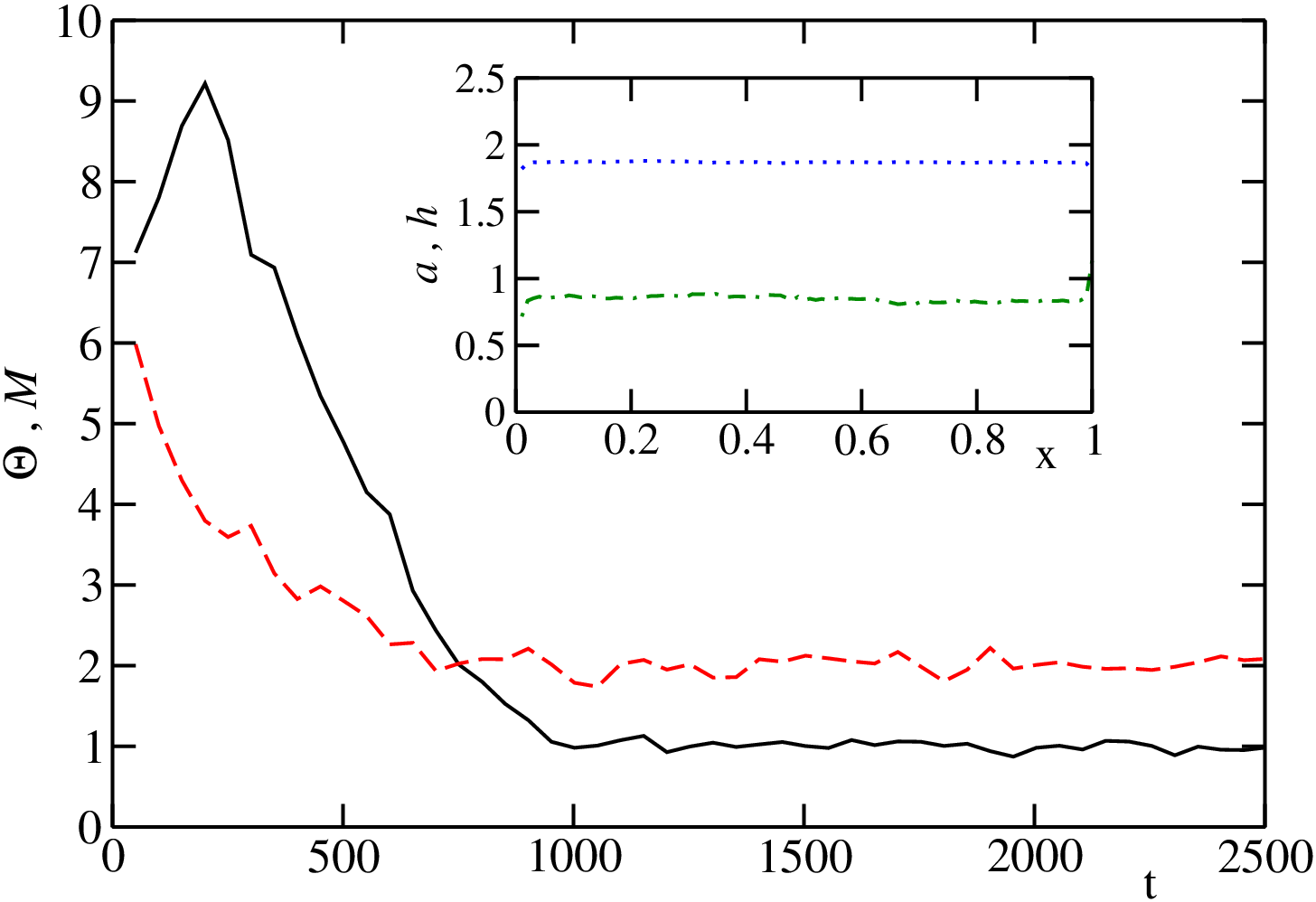}
\caption{Evolution of $\Theta$ (black solid line) and $\mathcal{M}$ (red dashed line)
 in a chain of $N=100$ sites in contact with two Langevin heat baths 
with $T_L=T_R=1$, $\mu_L=\mu_R=2$.
The initial state corresponds to $a=4$ and $h=15$. $\Theta$ and $\mathcal{M}$
are measured according to \cite{Iubini2012} and averaged over running windows
of 50 time units. In the inset: final spatial profiles of 
mass (blue dots) and energy (green dot-dashed line) densities as a function
of the rescaled site index $x=n/N$. 
}
\label{f:relax}
\end{center}
\end{figure}

The simple case of  a chain in contact
with two reservoirs, operating at the same temperature and chemical potential,
allows to test the Langevin scheme (\ref{olla2}). In figure~\ref{f:relax} we show a typical 
relaxation process towards an equilibrium state, characterized by the temperature and 
the chemical potential imposed by the reservoirs. The quantities $\Theta$ and $\mathcal{M}$
on the vertical axis denote the dynamical observables of the DNLS chain representing  its microcanonical
temperature and chemical potential 
respectively. Such quantities are complicated functions of the canonical variables, see  
\cite{Franzosi2011, Iubini2012} for details.
The inset in  figure~\ref{f:relax} shows that the asymptotic state reached
after the relaxation process is a genuine equilibrium state, corresponding to spatially 
homogeneous mass and energy densities. 

For the sake of completeness, we have also checked the equivalence between the
Langevin scheme and the MC reservoirs. As an example, in figure~\ref{f:ph_diag}
we show that the isochemical lines $\mu = 2$, obtained with the two approaches,
essentially coincide.

We conclude this section by observing that the scheme (\ref{ollac}) reduces, for
$T\to\infty$, to a standard Langevin formulation. Actually, in the large temperature 
limit, but finite mass- and energy-densities, it turns out that $\mu \to -\infty$ 
and $\gamma \to 0$. In this limit (\ref{olla2}) reduces to
\begin{equation}
i \dot z_1= -2|z_1|^2 z_1 -z_{2}
 -i\Gamma z_1+\sqrt{a_L \Gamma }\, \xi_1 \ \quad ,
 \label{langTinf}
\end{equation}
where $\Gamma=-\gamma \mu_L>0$ and $a_L = -T_L/\mu_L$ are finite quantities. 
Notice that $a_L$, which corresponds to the average mass density in the first
site, plays the role of an effective temperature.

\begin{figure}[h]
\begin{center}
\includegraphics[width=0.7\textwidth,clip]{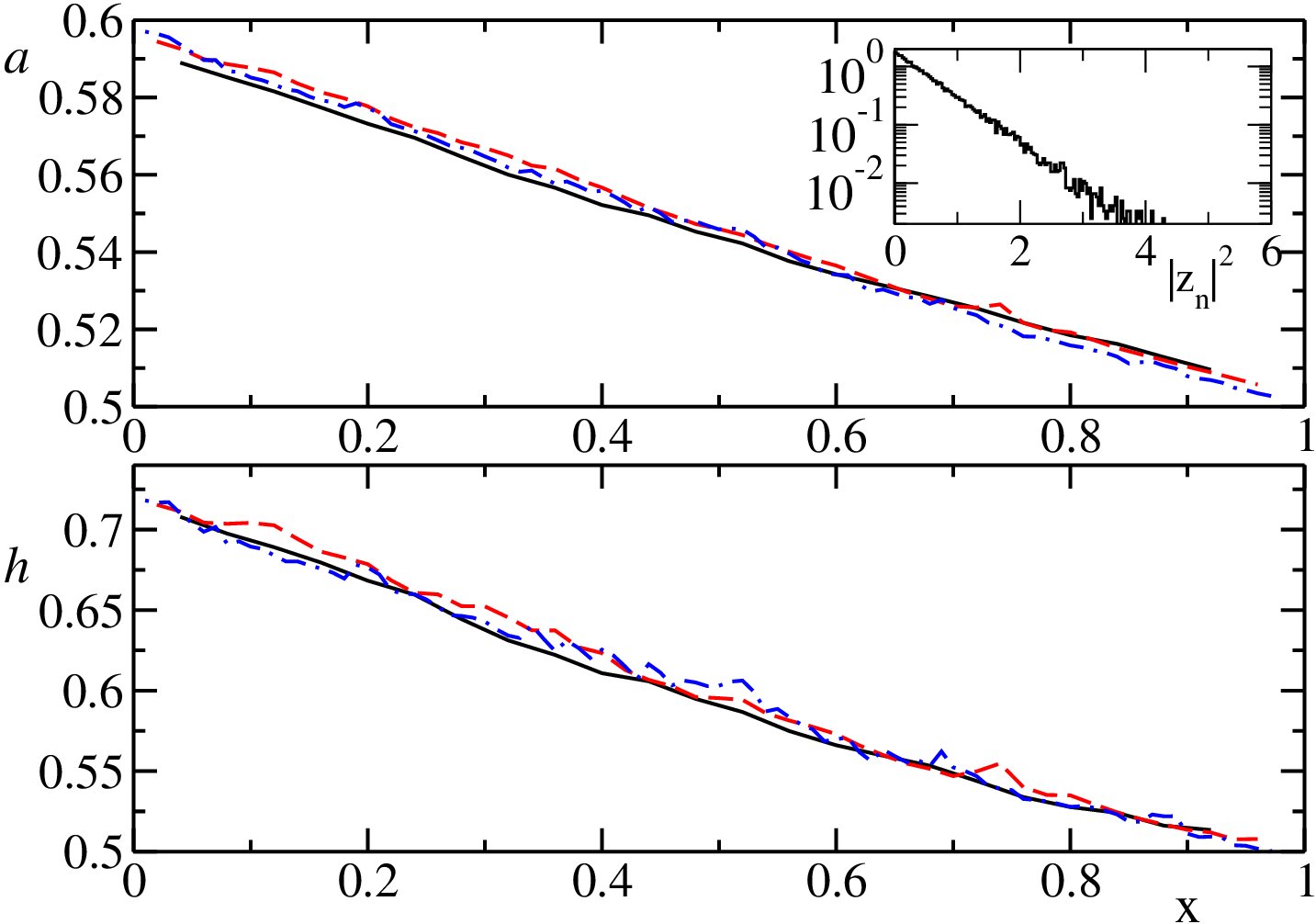}
\caption{Nonequilibrium profiles of mass density (upper panel) and 
energy density (lower panel) obtained  with the infinite temperature 
Langevin equation (\ref{langTinf}) and parameters $a_L=0.6$, $a_R=0.5$, $\Gamma=1$.
Black solid, red dashed and blue dot-dashed lines refer respectively to chain lengths
$N=25,50,100$. The profiles are almost linear and $h(x)=2a^2(x)$ along the chain, 
confirming that the thermostats act at $T=\infty$. The inset shows that
histogram of the local mass $|z_n|^2$ at $n=20$ is Poissonian, as expected.   
}
\label{f:prof}
\end{center}
\end{figure}

As a numerical check, we simulated eq.~(\ref{langTinf})
with $a_L > a_R$, i.e. in a nonequilibrium setting.  Even for
relatively short chains, the relation (\ref{tinfline}) is fulfilled at all points 
of the chain (see figure~\ref{f:prof}) meaning that local equilibrium holds. 
This is further confirmed by the shape of the distribution of the local mass, that is
Poissonian, as expected in the $T=\infty$ case \cite{Rumpf2004} 
(see the inset of figure~\ref{f:prof}). Establishement of local equilibrium
is not granted a priori, although it is known to generically occur in simulations of 
chains of nonlinear oscillators, even when transport is anomalous
\cite{LLP03}. This is nevertheless a peculiar case, as both temperature and 
chemical potentials are arbitrarily large.  However, it should be remarked that 
local equilibrium relations can be demonstrated to hold exactly only in very simple cases 
like for instance the Kipnis-Marchioro-Presutti model \cite{KMP82}. For the present model
(but also in other nonlinear chains) it is not at all
obvious that energy transfer among oscillators can be even roughly 
approximated by a Markovian process.

\section{The low-temperature and large mass density limits}
\label{sec:har}

In the previous section we have shown that the Langevin formulation of the DNLS 
thermodynamics provides a clear physical interpretation of the infinite-temperature limit. 
In a recent paper \cite{Iubini2012}, it has been found that the opposite, low-temperature, 
limit, as well as the case of large mass-densities, is also interesting for its nontrivial 
transport properties. In this section we show that the Langevin 
formulation can shed further light on such regime by revealing a
strong relationship with the XY model and thereby bridging a gap between two seemingly 
different classes of systems.

It is, first of all, convenient to introduce the following change of variables
\begin{equation}
{z}_n = \sqrt{a}(1 + \zeta_n)\exp[i(\omega t + \phi_n + n\pi) ] \quad ,
\label{ground}
\end{equation}
where $\omega= 2(a-1)$. By inserting (\ref{ground}) into (\ref{uno}), 
one finds that the new variables $\zeta_n$ and $\phi_n$ obey the dynamical equations 
\begin{eqnarray}\label{noapprox}
\hspace{-1.cm}\dot{\phi_n}(1+\zeta_n) &=& 2(1+2a)\zeta_n+2a(3\zeta_n^2+\zeta_n^3)-
(1+\zeta_{n+1})\cos(\phi_{n+1}-\phi_n)\nonumber  \\
\hspace{-2.cm}&&-(1+\zeta_{n-1})\cos(\phi_n-\phi_{n-1})+2 \\
 \dot{\zeta_n} &=& (1+\zeta_{n+1})\sin(\phi_{n+1}-\phi_n) -
(1+\zeta_{n-1})\sin(\phi_n-\phi_{n-1}) \quad .
\nonumber
\end{eqnarray} 
In this representation, the ground state (\ref{groundstate1}) is simply $\zeta_n = 0$, $\phi_n= 0$. 
Also, this is the optimal starting point to discuss two limit cases:
low-temperatures and large mass-densities.
The low-temperature dynamics can be studied by assuming $\zeta_n\ll 1$ and 
($\phi_n-\phi_{n-1})\ll 1$. In Appendix 2, we show that, in this a limit, the system is 
equivalent to a chain of harmonic oscillators with nearest- and next-to-nearest-neighbour interactions.
Furthermore, mass conservation of the original model maps into momentum conservation for the oscillators.
Such a correspondence is instructive, as it reveals a link with separable models, where a simple local definition
of the temperature can be given in terms of the kinetic energy. 

The large mass-density regime is studied under the assumption that $a\gg 1$ and $\zeta_n \ll 1$.
Here below we show that also in this approximation the Hamiltonian becomes separable. In fact,
the dynamical equations (\ref{noapprox}) reduce to leading order to
\begin{eqnarray}
\label{eqxy}
 \dot{\phi_n} &=& \lambda_n\\
 \dot{\lambda_n} &=& 4a \left[\sin(\phi_{n+1}-\phi_n)-\sin(\phi_n-\phi_{n-1}) \right]\nonumber
 \quad ,
\end{eqnarray}
where $\lambda_n=4a \zeta_n$. This system corresponds to a system of coupled rotors, 
i.e. a classical version of the XY model in one 
dimension \cite{Giardina99,Gendelman2000,Yang2005,Escande1994}. Its Hamiltonian reads 
\begin{equation}\label{HXY}
 \mathcal{H}_{XY}=\sum_n \frac{\lambda_n^2}{2}-\sum_n {4a\cos(\phi_{n+1}-\phi_n)} \quad ,
\end{equation}
where $\lambda_n$ and $\phi_n$ are a couple of conjugate action-angle variables, 
the former playing the role of the angular momentum. This analogy 
was already noticed (for the two-dimensional case) in \cite{Trombettoni2005}. 

At variance with the former case, we have not introduced any smallness hypothesis for $\phi_n-\phi_{n-1}$; 
as a result, some nonlinear terms are maintained and one can, thereby, explore large temperatures as well.
Having assumed that $\zeta_n \ll 1$, this regime can be called {\it phase} chaos.
It is also interesting to observe that in the large mass--density limit the invariance 
under global phase rotations of the DNLS transforms
into the invariance under a translation of the angles $\phi_n$. Accordingly,  the conservation of the
total mass $A$ transforms into the conservation of the total angular momentum 
$L=\sum_n \lambda_n$ (this can be easily verified by expanding expression (\ref{norm})).
Notice also that the low-temperature limit discussed in Appendix 2 is not fully contained into the 
large mass-density regime, as it includes the case of relatively small $a$-values.

Before passing to thermodynamic studies, it is necessary to clarify the range of validity 
of the XY model as an approximation of the DNLS one.
The condition $\zeta_n\ll 1$ implies $\lambda_n\ll a$, i.e. $T\ll a^2$, because
on average  $\lambda_n^2$ is equal to the temperature $T$. As we are exploring
the range of large $a$-values, one can conclude that, the larger $a$, the broader
the temperature range where the XY model provides an accurate description of the DNLS equation.
Before drawing this conclusion, it is, however, necessary to be more careful. In fact, the
presence of a finite conductivity in the XY model can be traced back to the existence
of (possibly infrequent) jumps of angle-differences across the sinusoidal potential barrier. 
In the context of Eq.~(\ref{HXY}), the height of this barrier is of the order of $a$,
which is smaller than the maximal acceptable energy $a^2$ (since  $a\ll a^2$, for $a\gg 1$).
Accordingly the XY model provides an accurate description also of the barrier jumps and,
more than that, the validity of the XY model extends to the high-temperature regime
(here ``high" means above $a$) characterized by frequent jumps.

\subsection{Thermostatted chain}

Here, we examine how to describe the bath dynamics within the XY approximation.
Let us study the simple setup of a DNLS chain in contact with an external Langevin reservoir 
at the first site.
In the low-temperature limit, i.e. close to the ground state,
equation~(\ref{ollac}) specializes to
\begin{equation}\label {ollaLT}
\hspace{-2.cm}  i\dot z_1=-2|z_1|^2 z_1- z_2 -i\gamma \left[ 2|z_1|^2 z_1+z_2 -
 2(a-1) z_1 -\delta\mu \, z_1 \right]+
 \sqrt{\gamma T_L} \, \xi_1 \, .
\end{equation}
We have consistently assumed the chemical potential to be a perturbation of the 
ground-state value, i.e. $\mu = \omega + \delta \mu$ with $\delta \mu \ll \omega$. For $a \gg 1$,  (\ref{ollaLT}) 
transforms, to leading order in $a$, 
into a Langevin equation for the XY model with suitable dissipation and 
fluctuation terms
\begin{eqnarray}
\dot{\phi_1} &=&4a\zeta_1\\
\dot{\zeta_1} &=& \sin(\phi_{2}-\phi_1) -
\gamma\left(4a \zeta_1-\delta\mu\right) +
\sqrt{\frac{\gamma T_L}{a}} \, \xi_1 \nonumber \quad .
\end{eqnarray}
By then introducing the momenta $\lambda_n$, equation ({\ref{eqxy}), and the 
rescaled dissipation parameter $\gamma' =4a \gamma $, one finally obtains
\begin{eqnarray}
\label{XY+bath}
\dot{\phi_1} &=&\lambda_1\\
\dot{\lambda_1} &=& 4a \sin(\phi_{2}-\phi_1) -
\gamma'\left(\lambda_1-\delta\mu\right) +
\sqrt{4\gamma' T_L} \, \xi_1\nonumber \quad .
\end{eqnarray}
These equations describe a rotor chain in contact in the first site with a reservoir
at temperature $2T_L$ and constant torque $\gamma' \delta \mu$. 

This derivation provides an interesting interpretation of the DNLS chemical potential 
in the large mass-density limit: at equilibrium (\ref{XY+bath}) is expected 
to sample microstates compatible with the 
grandcanonical measure  $\exp{\{-\beta[\mathcal{H}_{XY}- \delta\mu L]\}}$,
where $L= \sum_n \lambda_n$ is the total angular momentum of the
XY chain and $\delta\mu$ can be interpreted as the average angular velocity
of the rotors.

\subsection{Nonequilibrium conditions}

The above formulation
can be straightforwardly extended to nonequilibrium setups, where 
$T_R \neq T_L$ and $\mu_R \neq \mu_L$. 
In this case, it is convenient to perform the XY approximation with
respect to a ground state that corresponds to the average chemical potential 
$(\mu_R+\mu_L)/2$. Accordingly, the resulting XY chain turns out to be 
forced by \textit{opposite external torques} $\pm \gamma'\delta\mu$, where now
$\delta\mu=(\mu_R-\mu_L)/2$.

The observables of major interest in the nonequilibrium context are the 
fluxes of the conserved quantities.
The continuity equations for mass and energy densities of the DNLS model 
allow determining their explicit expressions 
\begin{eqnarray} 
 j_n^a&=&i(z_nz_{n-1}^*-z_n^*z_{n-1})\\
\label{XYhfluxIN}
 j_n^h&=&\dot z_nz_{n-1}^*+\dot z_n^*z_{n-1} \quad .
\end{eqnarray}
In the large mass-density limit, the leading terms read
\begin{eqnarray} \label{XYfluxes}
j_n^a&=&-2a\sin(\phi_{n+1}-\phi_n ) \\
\label{XYfluxes2}
j_n^h&=& - 2 \omega \, a\sin(\phi_{n+1}-\phi_n)- 4 a\dot\phi_n\sin(\phi_{n+1}-\phi_n) \quad ,
\end{eqnarray}
where the variables $\zeta_n$ have been expressed in terms of $\dot \phi_n$.  Notice that 
the simple symmetric form of the second equation has been obtained by adding to (\ref{XYhfluxIN}) the 
quantity $-a (\dot\phi_n - \dot\phi_{n-1}) \sin (\phi_n - \phi_{n-1})$, 
whose average is zero in a stationary state. Eq.~(\ref{XYfluxes}) is just the 
momentum flux of the XY model, i.e. the local force. 
The term proportional to $\dot \phi_n$ in Eq.~(\ref{XYfluxes2}) has the typical structure of the energy
flux in the XY model: it is nonzero only at finite temperatures. 
The first term is a coherent contribution that results from the fact that the 
oscillators rotate with an average common frequency $\omega$: it survives
in the zero-temperature limit, when $j^h_n = \mu j^a_n$, so that the heat current $j^h_n-\mu j^a_n=0$,
i.e. there is no heat transport and no entropy production.
At low temperatures, Eqs.~(\ref{XYfluxes},\ref{XYfluxes2}) describe the Josephson effect, 
where the chain amounts to a single junction in between two superfluids \cite{Smerzi2002}.
The mass current is proportional to the phase gradient and is independent of the system length $N$,
i.e. it provides a ballistic contribution.

\section{Comparison with numerical simulations}
\label{sec:sim}

As mentioned above, an important difference between the DNLS equation and oscillator chains
(like the Fermi-Pasta-Ulam or Klein-Gordon models) is that its Hamiltonian
is not the sum of kinetic and potential energies. Therefore, it is
not obvious how to directly monitor the temperature $T$ and the chemical
potential $\mu$ in actual simulations. The only general approach we are 
aware of is based on non-local microcanonical expressions $\Theta$ and $\mathcal{M}$ 
\cite{Franzosi2011b} which, unfortunately are rather awkward to
compute in practice (see \cite{Iubini2012} for details). 
The perturbative analysis of the low-temperature limit and the correspondence with
the XY model show that the temperature $T$ and the chemical potential $\mu$ can be 
determined in terms of local variables. This is quite a relevant observation as it may
be used to simplify the definition of such thermodynamic quantities.
In the following we explore the range of validity of such definitions.

From (\ref{HXY}) it is straightforward to define a kinetic temperature as the fluctuations 
of the momentum $\lambda_n$ with respect to its average value $\langle\lambda_n\rangle$,
\begin{equation}
T_{XY} = \langle \lambda_n^2 \rangle - \langle \lambda_n \rangle^2 \quad .
\label{tempb}
\end{equation}
If we compare the  stochastic term in (\ref{XY+bath})
with the one imposed by the fluctuation-dissipation theorem and commonly 
used in the Langevin equation for oscillator models, $\sqrt{2\gamma T}$ (see \cite{LLP03,DHARREV}), 
we can conclude that our definitions imply $T_{XY}=2T$
(the factor 2 is just a consequence of the choice of the transformation of variables).

In figure \ref{fig:txy} we compare the general microcanonical definition of 
temperature for the DNLS model $\Theta$,  defined as in \cite{Iubini2012},  with 
$T_{XY}$ for an equilibrium setting, i.e. external reservoirs  at equal temperature and chemical
potential; $T_{XY}$ is computed by evaluating, in the same simulation, the average 
of the $\zeta_n^2$  defined in (\ref{ground}). 
The data clearly show that, by increasing
the chemical potential $\mu$ (i.e., by increasing $a$, since $\mu = 2(a-1)$), the range of values in which the two temperatures
coincide increases, as expected from the previous considerations.
On the other hand,
outside the limits of validity of the XY approximation discussed in section \ref{sec:har}, $\Theta$ and $T_{XY}$
can be strongly different from one another. In such regimes, $\Theta$ is the only valid 
definition of temperature. In the inset of figure~\ref{fig:txy} we show that the curves obtained
for different values of $\mu$ quite well collapse onto each other by rescaling both 
$T_{XY}$ and $\Theta$ by the factor $a^{-2}$. This implies that the range of validity of the
correspondence between these two temperatures increases proportionally to $a^2$.

\begin{figure}[h]
\begin{center}
\includegraphics[width=10 cm,clip]{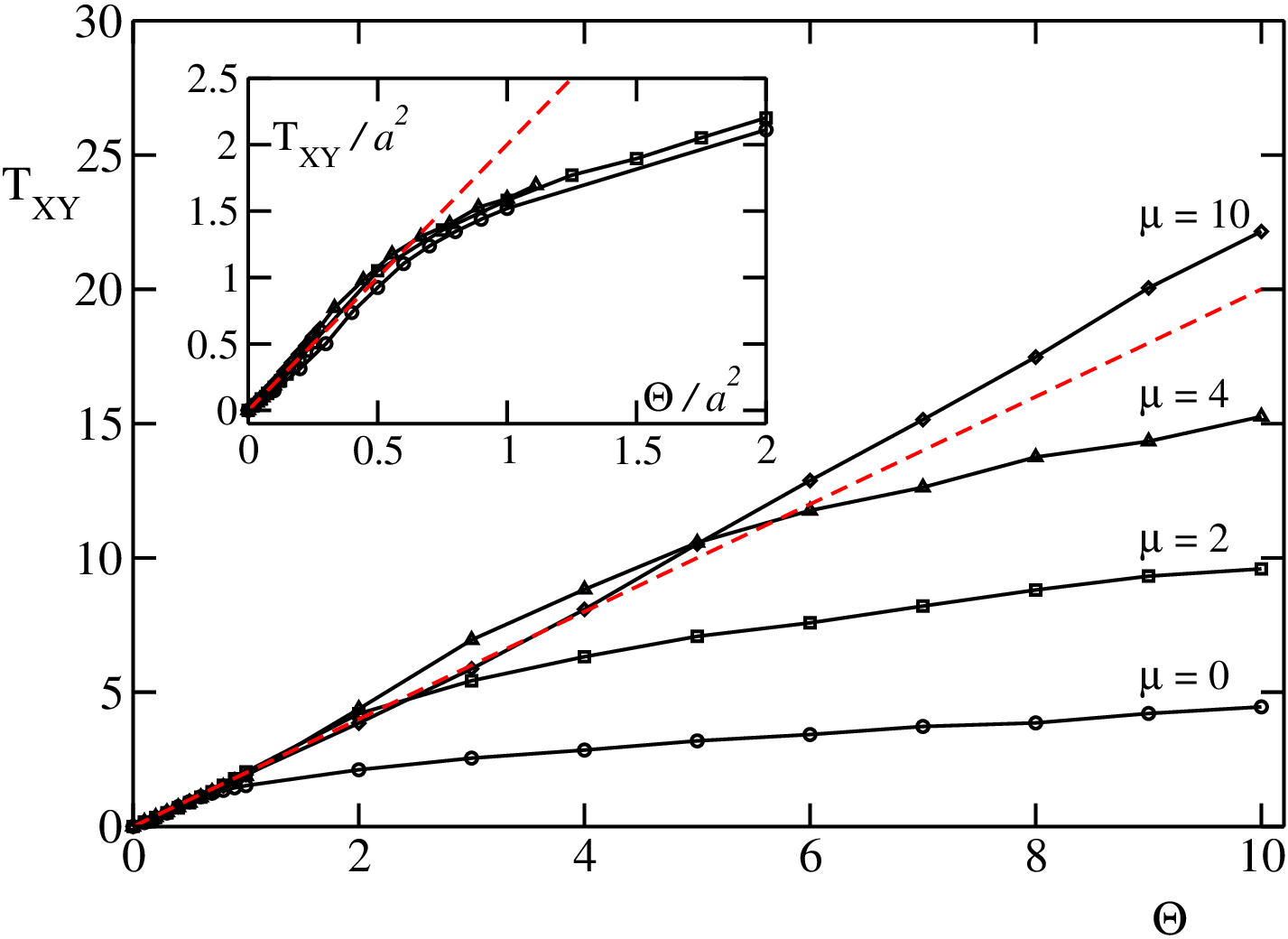}
\caption{Comparison of the XY kinetic temperature $T_{XY}$ with 
 $\Theta$ for different values of the chemical potential $\mu$.
The dashed line corresponds to $T_{XY}=2\Theta$, which should 
hold in the limit of large $\mu$, where the XY approximation is valid.
The inset shows the same curves in the rescaled units $T_{XY}/a^2$ and $\Theta/a^2$.
Simulations are performed using Langevin heat baths coupled 
at the boundaries of a DNLS chain with $N=50$. 
$T_{XY}$ and $\Theta$ are measured on a subchain of $30$ lattice sites.
{\bf }}
\label{fig:txy}
\end{center}
\end{figure}

\begin{figure}[h]
\begin{center}
\includegraphics[width=10 cm,clip]{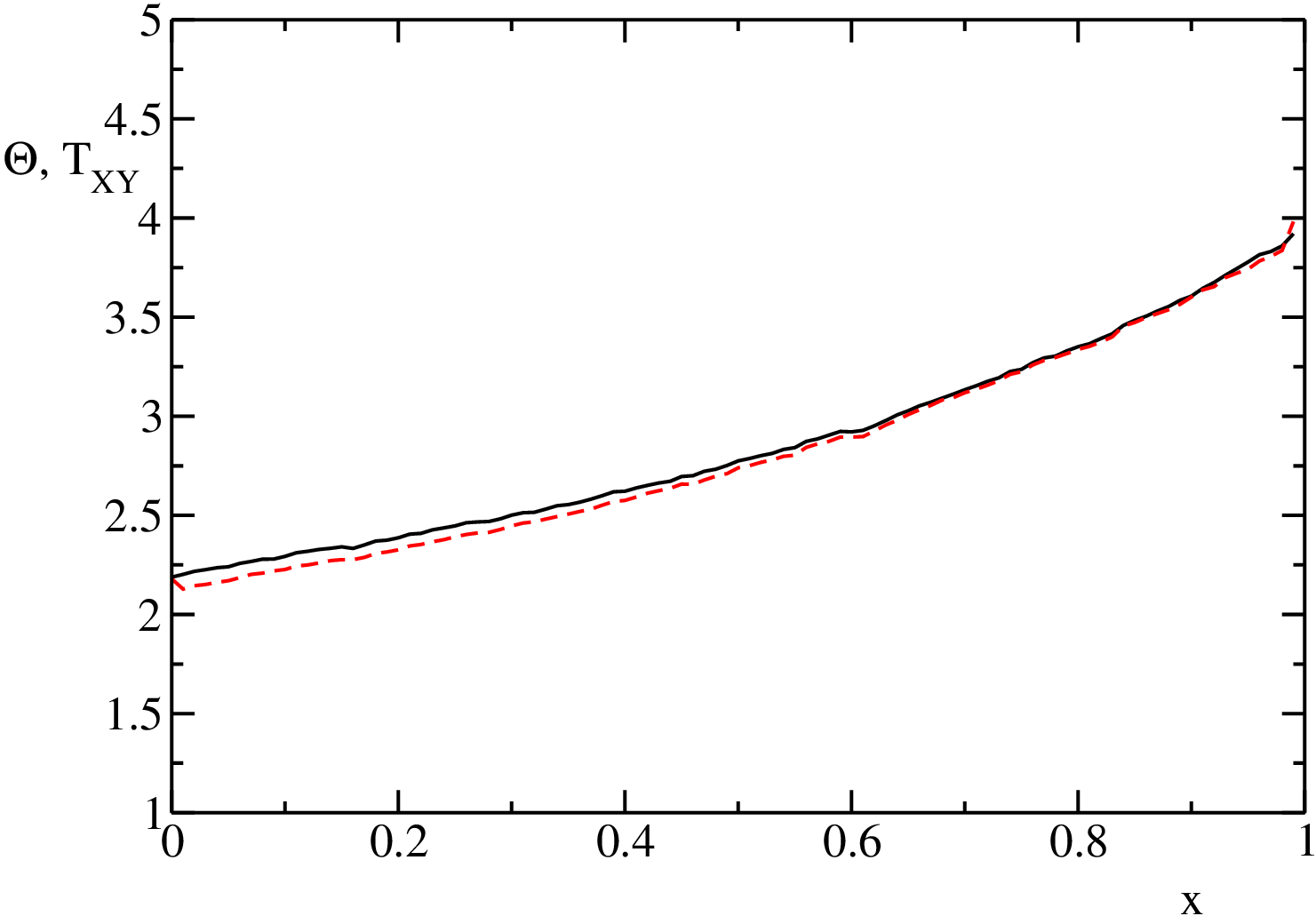}
\caption{Comparison of the XY kinetic temperature profile $T_{XY}$ (red dashed line) with 
 $\Theta$ \cite{Iubini2012} (black solid line) in a nonequilibrium steady state. 
Simulations are performed  using Langevin
heat baths with parameters ($T_L=2$, $\mu_L=9.6$) and ($T_R=4$, $\mu_R=10.4$), 
coupled to a DNLS chain with $N=1000$.}
\label{fig:NNeqT}
\end{center}
\end{figure}

 
Finally, we have tested the validity of the XY approximation in a non-equilibrium 
stationary regime. In the simplest case  one can impose two heat baths at different
temperatures, $T_R$ and $T_L$, and with the same chemical potential $\mu$ acting at the chain boundaries.   
If one chooses the value of $\mu$ in such a way that both temperatures are 
smaller than $a^2$ (see figure \ref{fig:txy}), one obtains temperature profiles very close to each other
(data not reported). This scenario is maintained also if a chemical potential gradient is applied,
provided both $\mu_R$ and $\mu_L$ are large enough to make the previous condition hold,
while $\delta \mu = (\mu_R - \mu_L)/2$ is smaller than the average chemical potential
$(\mu_R + \mu_L)/2$ (see sections 4.1 and 4.2 ).
In fact,  figure~\ref{fig:NNeqT} exhibits a nice agreement between the 
two temperature profiles. A further test of the validity
of XY model is presented in Table \ref{tab:fluxes} where the full DNLS fluxes
are compared with the ones reconstructed through the XY approximation 
(see Eqs.~(\ref{XYfluxes}) and~(\ref{XYfluxes2}) ~).
\begin{table}[h]
 \begin{center}
  \begin{tabular}{|c|c|c|}
   \hline
  & DNLS & XY\\
  \hline
  $j^a$& -0.234& -0.208\\
  $j^h$& -1.40&  -1.28\\
  \hline
  \end{tabular}
 \end{center}
\caption{Comparison of the exact DNLS fluxes (first column) and the ones reconstructed
by means of eqs.(\ref{XYfluxes}) (second column) for the nonequilibrium profile described
in the caption of figure~\ref{fig:NNeqT}.}
\label{tab:fluxes}
\end{table}   

\section{Conclusions}
\label{sec:end}

In this paper we have introduced Langevin heat baths
which are able to control both the temperature and the chemical potential
in a DNLS model. Numerical simulations indicate that such 
scheme is simple and practical enough to study finite-temperature 
DNLS dynamics in both equilibrium and nonequilibrium conditions.


In the low-temperature and large mass-density limit we have approximated 
the DNLS dynamics in terms of an effective XY model.
This allows a clear understanding of the DNLS dynamics, especially in a nonequilibrium setting. 
We have indeed shown that the effect of thermal baths 
(which are able to control the chemical potential besides the temperature) acting 
at the boundaries, is equivalent to an applied torque plus thermal fluctuations. 
This description allows to give a dynamical interpretation 
of the chemical potential as well as of the action of thermal baths 
as means to fix locally the average angular velocities.
The corresponding energy flux turns out to be the sum of two different contributions,
one due to the phase gradient associated with the torque,
the other due to angular-velocity fluctuations. As a consequence,
transport in this region has an almost ballistic component,
and a diffusive one associated with the XY dynamics which is known
to be a normal heat conductor \cite{Giardina99,Gendelman2000}.
This accounts for several previous computations of transport 
coefficients \cite{Iubini2012}.

Another remarkable result, is that the relationship with the XY model provides a 
simple prescription 
for computing the temperature in the simulations. This last issue is of major 
importance for non-standard Hamiltonians like the DNLS one, where kinetic and 
potential energies are not separated. Indeed, the XY approximation 
allows introducing the simple kinetic expression $T_{XY}$ for the temperature,
that can safely approximate the microcanonical one $\Theta$.
This is of practical importance, considering that the microscopic 
definitions of $T$ and $\mu$ are pretty much involved for a
non separable Hamiltonian, like the DNLS one \cite{Franzosi2011b}.

Altogether, starting from the Langevin approach we have achieved a fairly clear 
physical interpretation of the action of thermal baths as means to fix locally the 
average angular velocities
and kinetic energies of the oscillators. 
In the framework of the nonequilibrium XY model this means that one can explore 
more general nonequilibrium states by applying not only temperature but
also mechanical (torque) gradients. This possibility has not received much 
attention in the literature. To our knowledge, only reference \cite{Iacobucci2011}
treats the joint effect of thermal and mechanical gradients (see also 
the nonequilibrium studies in \cite{Eleftheriou2005} that however refer to
the case without external torque and noise).

Some of the numerical results presented above are possible starting 
points for rigorous investigations. For instance, the 
evidence of local equilibrium reported in section \ref{sec:lang}
and the possible approximate description in terms of 
stochastic models \cite{KMP82} could be a challenging issue
for mathematical studies.

Another possible extension of the present work would be to consider 
the DNLS model on two-dimensional lattices. In this case, the 
correspondence with the XY model would predict the 
possibility of observing the transition from normal to 
anomalous behavior of transport coefficients 
at the Kosterlitz-Thouless-Berezhinskii transition
\cite{Leoncini1998,Delfini2005}.

\ack
We thank S. Olla and Y. Dubi for fruitful discussions.

\section*{Appendix 1: Derivation of the Langevin equation}

In this appendix we derive Eq.~(\ref{ollac}) by following the system-bath 
coupling approach \cite{Zwanzig2001}.
In analogy with what done for harmonic lattices \cite{Dhar2006}, we
consider a complex oscillator, described by the dynamical variable $z$, linearly
coupled with a bath of independent, complex harmonic oscillators described by the 
Hamiltonian 
\begin{equation}\label{HB}
 H_B= \sum_\nu \{\omega_\nu^a |a_\nu|^2+\omega_\nu^b |b_\nu|^2+
\left[K_\nu^*z(a_\nu^*+b_\nu)+c.c.\right]\}
\quad ,
\end{equation}
where we have introduced two different species of oscillators, corresponding to the two sets of frequencies 
$\omega_\nu^a$ and $\omega_\nu^b$, while $K_\nu$ are the bath-system coupling constants. 
Moreover, the variables  $(a_\nu, ia^*_\nu)$ and $(b_\nu, ib^*_\nu)$ are independent canonically
conjugate coordinates, satisfying the following Poisson brackets
\begin{eqnarray}
\{ia_{\nu}^*,a_{\nu'}\} = \{ib_{\nu}^*,b_{\nu'}\} = \delta_{\nu,\nu'}\\
\{a_\nu,a_{\nu'}\} = \{b_\nu,b_{\nu'}\}= \{a_\nu,b_{\nu'}\} = \{a_\nu,ib_{\nu'}^*\} = 0\nonumber \quad .
\end{eqnarray}
In order to preserve the global symmetry of the  system with respect to phase transformations, we impose a second
conservation law,
\begin{equation}
A_B=\sum_\nu (|a_\nu|^2-|b_\nu|^2)\quad.
\end{equation}
The function $A_B$ is the generator of phase transformations of the bath variables. It is easy to
verify that the transformation generated by $A_B+|z|^2$,
\begin{eqnarray*}
 a_\nu(s)&=&e^{is}a_\nu(0)\\
 b_\nu(s)&=&e^{-is}b_\nu(0)\\
 z(s)&=&e^{is}z(0)\quad ,
\end{eqnarray*}
leaves the Hamiltonian $H_B$ invariant.
An example of heat bath satisfying these conditions  is given by a complex d'Alembert
 equation, $\Box\phi(x,t)=0$, for which the quantity $A_B$ represents the total (conserved) charge
of the field. 
The equations  of motion generated by (\ref{HB}) 
are
\begin{eqnarray*}
&& i\dot a_\nu= -\omega_\nu^a\, a_\nu -K_\nu^* z\\
&& i\dot b_\nu= -\omega_\nu^b\, b_\nu-K_\nu z^*\\
&& i\dot z = f(z) -\sum_\nu K_\nu (a_\nu+b_\nu^*) \quad ,
\end{eqnarray*}
where $f(z)$ accounts for the deterministic part of the evolution of $z$, not included in $H_B$.
The first two equations can be formally solved, yielding
\begin{eqnarray*}
&& a_\nu(t)=a_\nu(0)\,e^{i\omega_\nu^a t}+iK_\nu^*\int_{0}^t \,e^{i\omega_\nu^a(t-t')} z(t')\, dt'\\
&& b_\nu(t)=b_\nu(0)\,e^{i\omega_\nu^b t}+iK_\nu\int_{0}^t \,e^{i\omega_\nu^b(t-t')} z^*(t')\, dt' \, .
\end{eqnarray*}
By then substituting into the equation for $z$, we obtain
\begin{eqnarray*}
&&i\dot z = f(z) -i \int_{0}^t G(t-s) z(s) ds + F(t)  \quad ,
\end{eqnarray*}
where the noise term $F(t)$ and the dissipation function $G(t)$ are defined as 
\begin{eqnarray}
\label{appF}
&& F(t)=-\sum_\nu K_\nu \left[a_\nu(0)e\,^{i\omega_\nu^a t}+b_\nu^*(0)\, e^{-i\omega_\nu^b t}\right] \\
\label{appG}
&&G(t) = \sum_\nu |K_\nu|^2  \left[e^{i\omega_\nu^a t}- e^{-i \omega_\nu^b t}  \right]\quad .
\end{eqnarray}
By now imposing a grandcanonical equilibrium distribution $P\sim \exp[-\beta (H_B-\mu A_B)]$ 
for the bath of oscillators (where $\beta=1/T$ is the inverse temperature) \cite{Rumpf2004},
we find that the correlation functions of $F(t)$ read
\begin{eqnarray}
\hspace{-2.cm}\langle F(t)F(t')\rangle &=& \langle F^*(t)F^*(t')\rangle = 0 \nonumber \\
\hspace{-2.cm} \langle F(t)F^*(t')\rangle &=&\sum_\nu |K_\nu|^2 
\left[ e^{i\omega_\nu^a(t-t')}\, \langle\, |a_\nu(0)|^2\, \rangle
+ e^{-i\omega_\nu^b(t-t')}\,\langle\,| b_\nu(0)|^2\,\rangle\right]= \nonumber\\
\label{appF2}
\hspace{-2,cm}&& \sum_\nu |K_\nu|^2 \left[ \frac{e^{i\omega_\nu^a(t-t')}}{\beta(\omega_\nu^a-\mu)}+
 \frac{e^{-i\omega_\nu^b(t-t')}}{\beta(\omega_\nu^b+\mu)} \right] \, ,
\end{eqnarray}
where, in order to have positive definite statistical weights, we have also to assume  
$\omega_\nu^a> \mu$ and $\omega_\nu^b> -\mu$. 
In the thermodynamic limit the sums over the index $\nu$ in (\ref{appG}) can be replaced 
by  integrals. Accordingly, we can rewrite Eq.~(\ref{appG}) in the form
\begin{equation}
G(t) = \int_{\mu}^{+\infty}\, d\omega \,G^a(\omega)\, e^{i\omega t} -
\int_{-\mu}^{+\infty}\, d\omega\, G^b(\omega)\, e^{-i\omega t}  \quad ,
\end{equation}
where $G^{a,b}(\omega)=\rho^{a,b}(\omega) |K(\omega)|^2$ are two positive definite functions
and $\rho^{a,b}(\omega)$ the corresponding density of states that we assume to be smooth
functions.
By following the same approach, Eq.~(\ref{appF2}) writes  
\begin{equation}
\hspace{-2.cm}\langle F(t)F^*(t')\rangle 
=\int_\mu^{+\infty} d\omega\,\frac{G^a(\omega)e^{i \omega (t-t')}}{\beta(\omega -\mu)}
+\int_{-\mu}^{+\infty} d\omega\, \frac{G^b(\omega)e^{-i \omega (t-t')}}{\beta(\omega +\mu)} \quad ,
\label{fdt}
\end{equation}
which is a kind of fluctuation-dissipation theorem \cite{Dhar2006} where the Bose-Einstein
distribution has been replaced by the Rayleigh-Jeans one. 

The corresponding generalized Langevin equation is not very practical, since it is non Markovian.
We have nevertheless the freedom to choose the coupling and the density of states of the bath. 
The spectral properties of the process $F$ strongly depend on the 
behaviour of $G$ close to the ground state and may also display long-range 
correlations. To understand this point, consider the example in which 
$G^{a,b}(\omega)=\gamma$. This choice yields a spectral density of $F(t)$ which
is logarithmically divergent close to the ground state frequency, thus defining
a non-stationary process.   
The simplest, nonsingular case is obtained by choosing 
\begin{equation*}
G^a(\omega)=\frac{\gamma}{2\pi}(\omega-\mu),\quad
G^b(\omega)=\frac{\gamma}{2\pi}(\omega+\mu)\quad .
\end{equation*}
In this case $F(t)$ becomes a complex white noise
\begin{equation*}
 \langle F(t)F^*(t')\rangle = \frac{\gamma}{\beta} \delta (t-t')\quad ,
\end{equation*}
while the dissipation function is
\begin{equation}
 G(t)=\frac{\gamma}{2\pi} \int_{-\infty}^{+\infty} d\omega \, (\omega -\mu) e^{i\omega t} =
 -\gamma \left[i \frac{d}{dt} \delta(t)+\mu \delta(t)  \right]\quad .
\end{equation}
The full dissipation term is therefore
\begin{equation}
 -i \int_{0}^{t} G(t-s) z(s)\, ds =-\gamma \dot z(t) +i\gamma\mu z(t) \quad ,
\end{equation}
and the resulting Langevin equation corresponds to a noisy, driven, complex
 Ginzburg-Landau equation 
\begin{equation}
 (i+\gamma)\dot z = f(z) +i \gamma \mu z  + F(t)\quad.
\end{equation}
In the weak coupling limit ($\gamma\ll 1)$, the equation can be further simplified.
By multiplying by $(1+i\gamma)$ and neglecting terms  $O(\gamma^{3/2})$, one obtains
\begin{equation}
 i\dot z = (1+i\gamma)f(z) +i\gamma\mu z + F(t) \quad,
\end{equation}
which has the same structure as Eq.~(\ref{ollac}).

\section*{Appendix 2: The low--temperature limit of the DNLS problem}
In this appendix we provide a low--temperature description of the DNLS equation in terms
of a harmonic model with separable Hamiltonian. In this limit, the solution of Eq.
 (\ref{noapprox}) is 
expected to be close to the homogeneous periodic motion of the ground-state solution 
(\ref{groundstate1}). Thus, we assume $\zeta_n\ll1$ and $(\phi_n-\phi_{n-1})\ll1$ 
and we expand Eq. (\ref{noapprox}) to linear order. As a 
result, we obtain
\begin{eqnarray}
\label{model}
\dot {\phi}_n &=& 4a\zeta_n +2\zeta_n - \zeta_{n+1}-\zeta_{n-1} \\
\dot {\zeta}_n &=& \phi_{n+1}-2\phi_n + \phi_{n-1}  \nonumber \quad .
\end{eqnarray}
If one now introduces the new variable
\begin{equation}
p_n = 4a\zeta_n +2\zeta_n -  \zeta_{n+1}-\zeta_{n-1} \quad .
\label{p}
\end{equation}
the Eqs.~(\ref{model}) can be re--written as
\begin{eqnarray}
\label{eq:osc}
\dot {\phi}_n &=& p_n  \\ 
\dot {p}_n &=& 4(1+a) ( \phi_{n+1}-2\phi_{n} +\phi_{n-1}) -\phi_{n+2}
  +2\phi_{n}-\phi_{n-2} \nonumber \quad .
\end{eqnarray}
These equations describe the dynamics of a chain of harmonic oscillators with 
nearest-neighbour and next-to-nearest-neighbour interaction. The corresponding 
Hamiltonian,
\begin{equation}
\label{hamil2}
\mathcal{H}_{h} =  \sum_n \left[\frac{1}{2} p_n^2 + 2(1+a) (\phi_{n+1}-\phi_{n})^2 - 
\frac{1}{2} (\phi_{n+2}-\phi_{n})^2 \right] \quad ,
\end{equation}
is, at leading order in $p_i$ and $(\phi_{n+1}-\phi_n)$, fully equivalent to 
that of the original DNLS equation. 
Its quadratic structure corresponds to a parabolic approximation around the minimum of the energy.
Moreover, the total mass-conservation law of
the DNLS maps onto the conservation of the total momentum $P =\sum p_n$. Accordingly, 
the Hamiltonian (\ref{hamil2}) is translationally invariant.
The normal modes, i.e. the plane-wave solutions of Eqs.~(\ref{eq:osc}), 
are the discrete analogs of the Bogoliubov modes (non-interacting phonons),
in the context of the physics of atomic condensates \cite{Smerzi2002}.

Passing to  thermodynamics, one interesting implication of the Hamiltonian structure 
in the low--temperature limit (\ref{hamil2}) is that one can naturally introduce a microscopic 
definition of temperature in terms of the momentum $p_n$, i.e.

\begin{equation}
T_h = J \left[\langle p_n^2 \rangle- \langle p_n \rangle^2  \right]
\label{tempa}
\end{equation}
where the proportionality constant $J$ is the Jacobian determinant of transformation
(\ref{p}), which must be included to allow for a meaningful comparison with the DNLS
model. In order to test the definition (\ref{tempa}),
we have measured $T_{h}$ by numerical simulations of the Langevin scheme 
defined in Eq. (\ref{ollac}) and in figure~\ref{f:Tharm} we have compared it with the 
temperature of the bath, $T_{B}\equiv T_L=T_R$, for different values of the mass density, $a$.
As expected, $T_{h}$ approaches  $T_{B}$ for increasing values of $a$. 
In fact, the larger is $a$, the smaller is the relative amplitude 
of the fluctuations with respect to the ground state.
From this analysis we therefore conclude that the harmonic temperature $T_h$ is 
a well defined thermodynamic observable in the low--temperature limit. Such a definition
is much simpler than the general microcanonical one, $\Theta$, defined in 
\cite{Iubini2012,Franzosi2011b}.

\begin{figure}[h]
\begin{center}
\includegraphics[width=0.7\textwidth,clip]{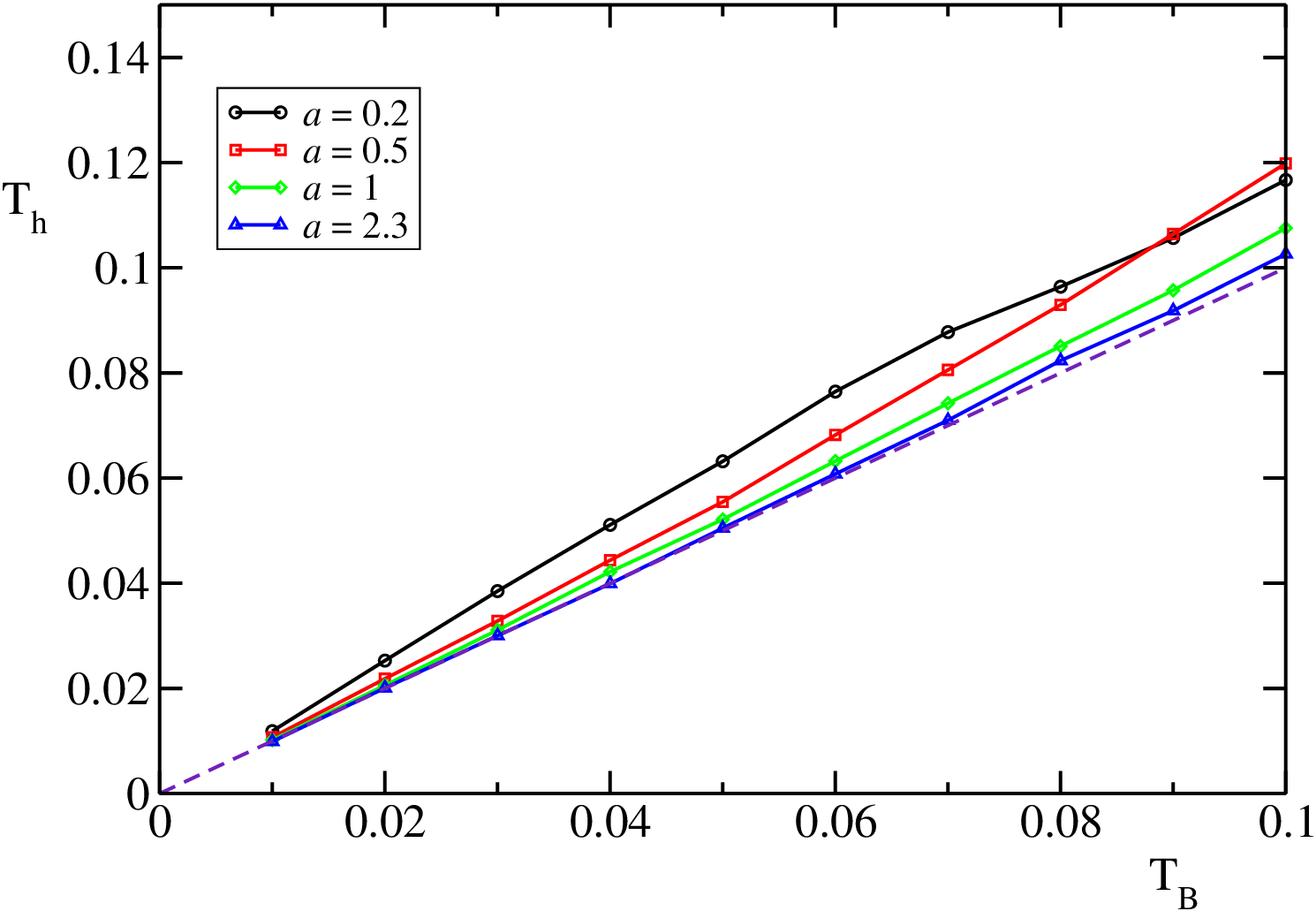}
\caption{Comparison of the harmonic temperature $T_{h}$, Eq.~(\ref{tempa}) with 
the reservoir temperature, $T_{B}$. The dashed line correspond to $T_{h}=T_{B}$.
The Langevin heat baths are coupled at the boundaries of a DNLS chain with $N=50$. 
$T_{h}$ is measured on a subchain of $30$ lattice sites to increase the statistics.
The values of the Jacobian determinant $J$, which are the product of the 
eigenvalues of a tridiagonal matrix, have been computed analytically in the 
$N\to \infty$ limit and are $J=5.95, 3.73, 2.91, 2.42$ for curves from
top to bottom.
}
\label{f:Tharm}
\end{center}
\end{figure}

For what concerns transport properties, the heat conductivity of  the harmonic
model (\ref{hamil2}) exhibits a divergence in the thermodynamic limit, as expected for any
integrable model (see \cite{RLL67,LLP03}). Such a conclusion is in contrast with 
previous non-equilibrium numerical studies of the DNLS model \cite{Iubini2012} that
have revealed a finite heat conductivity at finite temperatures.
This is clearly a consequence
of the presence of nonlinear terms which break the integrability of the dynamics. 
Their contribution  is taken into account in
section \ref{sec:har}, where we discuss
the  tight relationship with the one-dimensional XY model in the limit of large mass densities.
In this respect, it is useful to compare the harmonic hamiltonian $H_h$ (\ref{hamil2}) 
with the one corresponding to the XY model (\ref{HXY}). The former is valid in the low-temperature 
regime, while the latter applies for $a\gg1$ (and $T\ll a^2$). Accordingly, they reduce to one another 
for small $T$ and large $a$. In this limit, in fact,  the next--to--nearest-neighbour interaction 
in (\ref{hamil2}) is negligible for $a\gg1$ and, in the low--temperature limit, one can expand the cosine 
interaction in (\ref{HXY}) around zero.

\section*{References}

\providecommand{\newblock}{}

\end{document}